\documentclass[conference]{IEEEtran}
\IEEEoverridecommandlockouts
\usepackage{cite}
\usepackage{amsmath,amssymb,amsfonts}
\usepackage{algorithmic}
\usepackage{graphicx}
\usepackage{textcomp}
\usepackage{xcolor}
\usepackage{flushend}
\usepackage{tabularx}
\usepackage{caption}

\newcolumntype{b}{>{\hsize=.60\hsize}X}
\newcolumntype{s}{>{\hsize=.20\hsize}X}
\newcolumntype{t}{>{\hsize=.20\hsize}X}
\def\BibTeX{{\rm B\kern-.05em{\sc i\kern-.025em b}\kern-.08em
    T\kern-.1667em\lower.7ex\hbox{E}\kern-.125emX}}
\begin{document}

\title{SUPERNOVA: Automating Test Selection and Defect Prevention in AAA Video Games Using Risk Based Testing and Machine Learning
}

\author{\IEEEauthorblockN{Alexander Senchenko*}
\IEEEauthorblockA{\textit{Electronic Arts}\\
Vancouver, Canada \\
asenchenko@ea.com}
\and
\IEEEauthorblockN{Naomi Patterson*}
\IEEEauthorblockA{\textit{Electronic Arts}\\
Vancouver, Canada \\
npatterson@ea.com}
\and
\IEEEauthorblockN{Hamman Samuel}
\IEEEauthorblockA{\textit{Electronic Arts}\\
Waterloo, Canada \\
hsamuel@ea.com}
\and
\IEEEauthorblockN{Dan Ispir}
\IEEEauthorblockA{\textit{Electronic Arts}\\
Bucharest, Romania \\
dispir@ea.com}
}

\maketitle

\thispagestyle{plain}
\pagestyle{plain}

\let\thefootnote\relax\footnotetext{*These authors contributed equally to this work.}

\begin{abstract}
Testing video games is an increasingly difficult task as traditional methods fail to scale with growing software systems. Manual testing is a very labor-intensive process, and therefore quickly becomes cost prohibitive. Using scripts for automated testing is affordable, however scripts are ineffective in non-deterministic environments, and knowing when to run each test is another problem altogether. Manual testing and writing scripts make up the current industry standard and methodology for game testing, but the writing is on the wall for this practice. The modern game’s complexity, scope, and player expectations are rapidly increasing where quality control is a big portion of the production cost and delivery risk. Reducing this risk and making production happen is a big challenge for the industry currently. To keep production costs realistic up-to and after release, we are focusing on preventive quality assurance tactics alongside testing and data analysis automation. We present SUPERNOVA (Selection of tests and Universal defect Prevention in External Repositories for Novel Objective Verification of software Anomalies), a system responsible for test selection and defect prevention while also functioning as an automation hub. By integrating data analysis functionality with machine and deep learning capability, SUPERNOVA assists quality assurance testers in finding bugs and developers in reducing defects, which improves stability during the production cycle and keeps testing costs under control. The direct impact of this has been observed to be a reduction in 55\% or more testing hours for an undisclosed sports game title that has shipped, which was using these test selection optimizations. Furthermore, using risk scores generated by a semi-supervised machine learning model, we are able to detect with 71\% precision and 77\% recall the probability of a change-list being bug inducing, and provide a detailed breakdown of this inference to developers. These efforts improve workflow and reduce testing hours required on game titles in development.
\end{abstract}


\section{Introduction}

The purpose of quality assurance (QA) testing is to minimize defects in software products. Over time, as these software systems went from several thousand lines of code and few dependencies to millions and many, the cost of QA testing became exorbitantly more expensive while simultaneously being less effective compared to the past \cite{Munro2009}. This is because the size of software systems, and therefore the number of items that can be tested, is growing beyond what is reasonable to test uniformly. A common method to address this has QA testers select test cases manually based on some form of deductive reasoning, in an attempt to improve testing coverage and/or depth. Such mechanisms are subjective and do not produce sets of test cases that comprehensively assess the software in question. One possible solution for this is to automate the selection of test cases by using a probabilistic model to replace traditional efforts. This has the benefit of providing reasonably unbiased outcomes based on historical data which can be tuned over time. Another school of thought for improving software quality takes a more preventative approach, seeking to stop defects from entering software systems entirely, before they become bug inducing.

\bigskip

In this paper we present a solution which incorporates both of these concepts. On the test case selection front, given a software system with tests numbered $0...n$, our method chooses tests such that as $n$ approaches infinity, testing time remains constant while the efficacy of selection improves over time \cite{patent}. This method also identifies test scripts that have become irrelevant, stale or non-functional due to ongoing development, and removes them from the test execution scope. This test selection occurs in an end-to-end manner, as selections can be from any aspect of the software in question, allowing for an all-in-one solution. This results in an automated risk-based testing (RBT) system that formulates a risk assessment for automatic selection of both manual and automated QA tests. For defect prevention, given a commit with lines of code $1...n$, our method extracts features from this code data, the hierarchy of the software system, and even individual developer details to make an informed prediction on whether the code is bug inducing or not. This output is then presented to the developer with a breakdown of what features had the most impact on this decision, and they are then able to make an informed decision on whether to review their code. These systems are currently accomplished using probabilistic modeling and machine learning, with deep learning on the horizon as the next step for this project.

\bigskip

This new methodology is particularly relevant for the games industry, as video games have more uncertainty and potential points of failure than other software systems. Testing AAA games is a challenge that currently requires an extensive amount of manual testing efforts due to the significant scale of the testing scope. There is a time factor to this as well, because of the tight release cycles. The above factors explain why solutions that work for other tech companies do not necessarily translate into the games industry.

\bigskip

Although the focus of these methods are on testing video games and online services, they may be applied to any software application. We show that our approach is highly effective at reducing total work hours required to plan testing efforts and sufficiently test a software product. This marks a significant step in the process of fully automating software testing.

\section{Related Work}

There are many automated software testing systems currently available for consumer use. Katalon \cite{katalon} is an end-to-end software testing product that automates test generation for various development environments. They offer services for web, desktop, API and mobile environments. Likewise, other similar products such as Selenium \cite{selenium, selenium2} and TestComplete \cite{testcomplete, testcomplete2} offer comparable services in different environments. For many use cases, a solution such as these will be completely viable. However, for proprietary software environments, using third party solutions is not an effective way to conduct robust testing, as they will not be able to interface properly with internal software systems. 

\bigskip

Furthermore, in especially complicated systems such as game engines, manual test cases that can not be computer generated or evaluated will need to be run. For example, a manual test case can describe that a game tester navigate along the edges of a game level and attempt to break collision detection through the jump action. Furthermore, manual tests that seek to investigate visual and not logical anomalies often won't trigger any warnings or errors, relying solely on the game tester's ability to evaluate this themselves. There can be up to 40,000 of these manual test cases for a given game title. None of these test automating products are able to select, sort or choose predefined tests that report both automatically and manually input response data based on a mathematical model's output. Although there is research in selecting the data to use in creating a test case \cite{dataflow}, this only helps for test cases capable of being automatically generated. Other research seeks to link file dependencies with test cases such that they are only run them when the linked files are modified, rather than any file within the project \cite{testselection1, testselection2, testselection3}. These regression test selection approaches are very effective for test cases that are automatically executed, but are unable to accommodate manually executed test cases. Therefore there is a large gap in both research and the third party market for such a piece of software.

\bigskip

Alternatively, some software tools are used as a preventative measure for testing, rather than the typical reactive approach. These include Facebook's Infer \cite{infer}, which is a static code analyzer for several languages, capable of detecting a variety of errors such as null pointer exceptions. Another prominent example is Clever-Commit \cite{clevercommit}, a recent venture from Ubisoft and adopted by Mozilla, which incorporates machine learning to prevent bugs from entering production code at the commit level using code metrics, clone detection, and dependency analysis. This builds off the fundamental groundwork laid by what is commonly referred to as the SZZ algorithm, as well as Commit Guru \cite{szz, Kondo2020, Tabassum2020, Chen2014}. These are all interesting approaches to bug prevention, and were inspiration for the preventative efforts we present in this paper.
Lastly, the concept of RBT is not new, and has been previously researched with impressive results \cite{rbt, rbt2}. This prior work also served as inspiration for the test selection approach we used in SUPERNOVA.

\section{Methodology}

\subsection{Overview}

\begin{figure*}[h]
\centering
\includegraphics[width=1.0\textwidth]{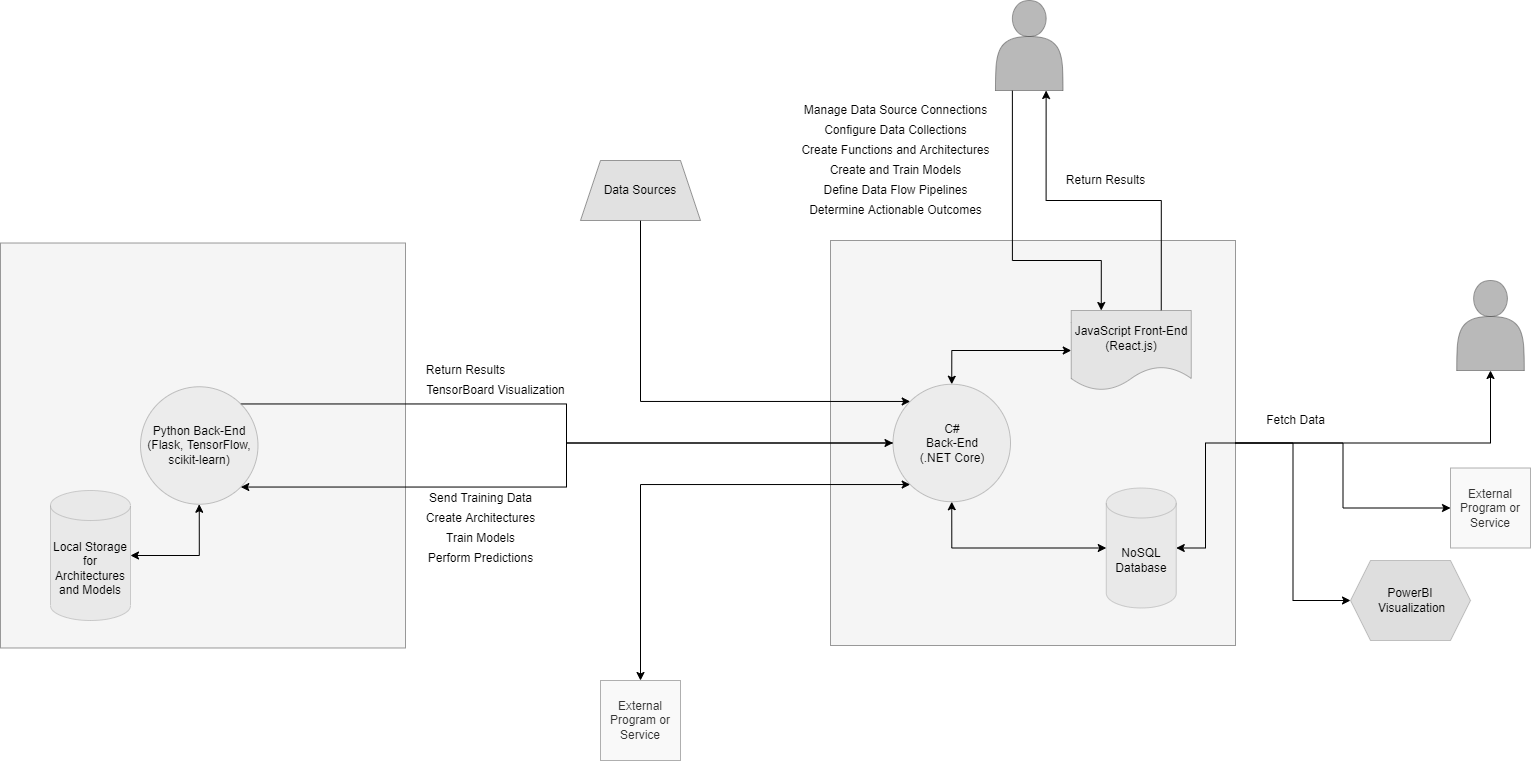}
\caption{High level overview of SUPERNOVA}
\label{SUPERNOVA}
\end{figure*}

SUPERNOVA captures end-to-end automation for data science based testing with mathematical models which automate workflows and drive decision making in QA testing for AAA games. There are two central components to this effort; an end-end automation tool for software testing, as well as a machine or deep learning model management and training service. SUPERNOVA is designed and internally distributed with the software as a service (SaaS) model, and is not used in conjunction with other end-to-end software testing frameworks. The general architecture of this system can be seen in Figure \ref{SUPERNOVA}.

\bigskip

\begin{figure}[h]
\centering
\includegraphics[width=0.45\textwidth]{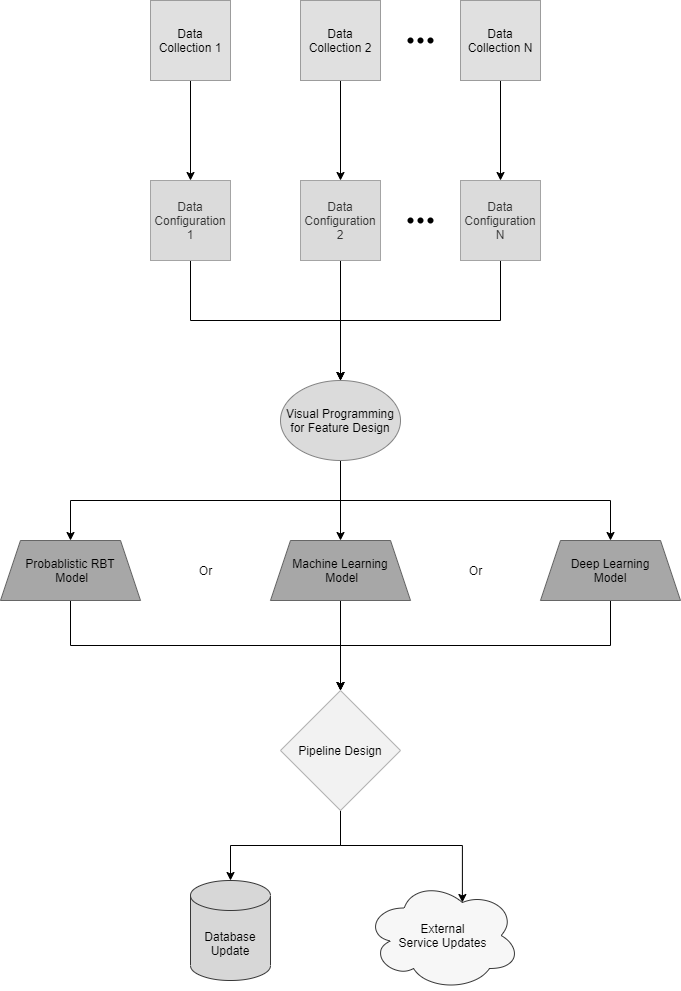}
\caption{Example of the SUPERNOVA workflow}
\label{workflow}
\end{figure}

Users interact with the SUPERNOVA user interface to manage data source connections, define data collections, design rules based data configurations, create mathematical functions through a visual programming interface, determine actionable outcomes, monitor deployed pipeline performance, pre-process training data, define deep learning architectures, train machine or deep learning models, and monitor model training performance. There is an outline of the general workflow for SUPERNOVA in Figure \ref{workflow}, and we will be presenting it's features in that same order.

\bigskip

Starting with data source connections and collections, an interface for all relevant input source structures is provided to the user within SUPERNOVA. For example, to retrieve QA data from a data source such as Jira, there are a number of expected parameters which are shown to the user. These include the URL instance, the project key, and JQL for key based filters such as severity or status of an issue. Other supported input structures include TestRail, DevTest, Git, Perforce, EA proprietary tools, SQL, NoSQL, and custom calls for unknown sources \cite{versionctrl}. This streamlines the otherwise tedious process of data collection by providing the building blocks to accomplish this efficiently.

\bigskip

Following input collection is data configuration, which allows for chains of fields to be used for indexing into sourced data structures. We developed our own framework that can interface with all supported data sources from the data collection phase and combine them into a unique proprietary format, which allows for efficient filtering, grouping, and searching. Users will recursively construct linked nodes that represent the linked fields within this format, after which all selected inputs will be ready for modeling and modification. An example of this would be to take a Jira Epic key from one data collection to index into another collection and retrieve data for that Epic not stored in the initial collection.

\bigskip

Once the desired data fields have been selected, they can be modified by SUPERNOVA's visual programming interface to define functions that create metrics to describe useful behaviours. These represent factors such as number of consecutive successful test runs, time since the test was initially created, frequency of bug discovery, etc. In the case of real number data fields, a manually tuned weight and activation function can be applied in order to provide further control. These heuristics can be as simple as a direct mapping, such as $f(x) = 2x$. However, the output from one function can also be directly sent as input to another. This allows for stacks of equations where each has the potential to branch depending on input, as conditional statements are allowed. 

\bigskip

\begin{figure}[h]
\centering
\includegraphics[width=0.45\textwidth]{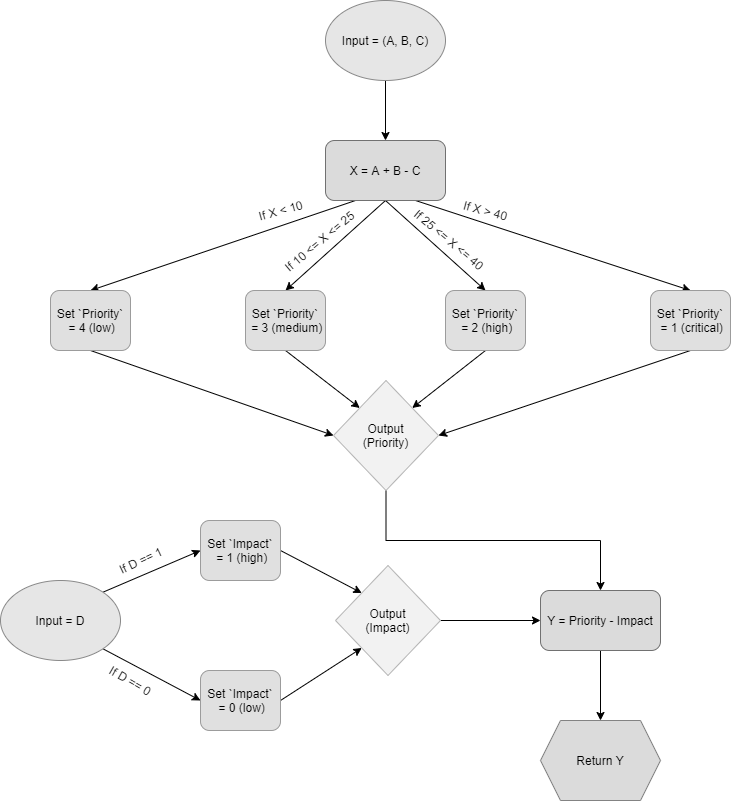}
\caption{Example subtree of a parent tree, used in one of EA's pipelines. Input A, B, C are weights or scores from one or more data sources. They are fed through a function producing an output (priority) depending on a conditional. Afterwards, this output is used by another function along with another input (impact) to return a final output, which is returned to the parent tree.}
\label{tree}
\end{figure}

Furthermore, return values can be specified, which allow recursion and other advanced methods when necessary. Therefore these equations can be described as decision trees, and an example of this is seen in Figure \ref{tree}. These trees are modular, as they use multiple predefined metrics combined together, and as such nodes or edges can easily be swapped out or changed.

\bigskip

This flexibility allows our data scientists to quickly build models for a wide variety of tasks that will thereafter function in an automated state. There are fundamentally three different methods to construct models. The first allows for simple mathematical formulae where functions can be created from desired inputs, using RBT. The risk exposure $R$ of a risk item $a$ is calculated as follows, and is in the range $[0, 100]$:

\begin{gather}
    R(a) = P(a) \cdot T(a) \cdot I(a)
\end{gather}

where $P$ is the probability, $I$ is the impact, and $T$ is the time factor of a risk item $a$, which are defined as:

\begin{gather}
    P(a) = \dfrac{\sum_{j=1}^{m} p_{j} \cdot w_{j}} {\sum_{j=1}^{m} w_{j}}
\end{gather}

\begin{gather}
    I(a) = \dfrac{\sum_{j=1}^{n} i_{j} \cdot w_{j}} {\sum_{j=1}^{n} w_{j}}
\end{gather}

\begin{gather}
    T(a) = \dfrac{\sum_{j=1}^{k} t_{j}} {k}
\end{gather}

$p_{j}$ are values in range $[0, 10]$ for $m$ probability criteria, $i_{j}$ are values in range $[0, 10]$ for $n$ impact criteria,  $t_{j}$ are values in range $[0, 1]$ for $k$ time criteria, and $w_{j}$ are weight values in range $[0, 1]$ for $m + n + k$ criteria. Note that the criteria are normalized to be within these ranges. Probability criteria are a set of metrics that represent the likelihood that the failure assigned to risk occurs. These are made up of technical criteria, such as code complexity, and inferred automatically. Impact criteria are a set of metrics that capture the cost or severity of failure if it occurs in operation. These are made up of business criteria, such as importance of a system's continued functionality, and inferred either automatically or manually. Time criteria are a set of metrics that adjust probability factors for risk items, with the idea being that risk items have a life cycle. These are made up of test criteria, such as testing history, and manually set to $1.0$ but automatically updated over time. For example, should testing history indicate that bugs are being frequently found in a game system, the criterion value stays close to $1.0$ for that system. Likewise, if few or no bugs are found, the value lowers to represent the decay of risk for that system. Therefore the time criteria can automatically stop tests that are stale, irrelevant or non-functional, and expose them to us. The weights for each of these factors are hand tuned by our data scientists for an individual project's needs and goals. The risk exposure is the final part of the tree, and returns a score which can then be used for sorting of tests and other applications.

\bigskip

The second method for model construction is machine learning, which uses JSON schema that interface directly with the scikit-learn library, allowing for the selection of any algorithm supported by the library alongside robust hyperparameter tuning for each. This is presented in the user interface as a card with dropdowns and text boxes allowing for all selection to be done without any coding required. A training target is also set from one or more of the configured inputs, which the model learns to reproduce on unseen data. Examples of this include whether a test found a bug, or if a commit is bug inducing. These selections are then sent to the machine learning endpoints for training or prediction. This is a key aspect of SUPERNOVA's appeal, as many of our QA and game testers do not know how to program, so providing an automated way of performing data science is very useful to EA at scale.

\bigskip

The third method for model construction is deep learning, which uses another JSON schema that interfaces with the TensorFlow library. This abstracts the typically complicated process of creating a deep learning architecture and training a model with it. This is simplified in SUPERNOVA's user interface with a drag and drop, node and edge system that also has no coding requirements. For architecture design, each layer in the neural network is a node that has edge(s) entering and exiting it. The edges entering represent inputs, and the edges exiting represent outputs. Within each node the type of layer, its activation function, initialization, and all other hyperparameters are set using a dropdown or text box. Furthermore, the structure of these nodes does not have to be linear. There is full support for functional (ie non-linear) type architectures as well. Once the architecture has been constructed using the interface, it is named and saved. Architectures can be easily shared to others if desired. Model training is then performed using a very similar visual card interface as from the machine learning method. Although SUPERNOVA is capable of any type of learning, since it is an interface into both scikit-learn and TensorFlow, the user interface is only configured to use supervised learning at this time, and as such one or more training targets must be selected. Lastly, TensorBoard is integrated for in depth model training monitoring, and TensorFlow pre-processing techniques can be automatically applied to any data sent to SUPERNOVA, such as the encoding of categorical variables. 

\bigskip

Compared to machine learning efforts, deep learning provides the ability to extract features from matrices automatically, without the need for us to do our own feature extraction or transform them into lower dimensional space. For example, this makes it possible to train a model directly on semantic code data, without needing to create features such as cyclomatic complexity, which simply reduce the source code into a numeric value.

\bigskip

By giving users the option to use a probabilistic model, a machine learning model, or a deep learning model, these choices end up creating inclusivity as they represent the corresponding skill levels of beginner, intermediate and advanced. This allows SUPERNOVA to have both depth and breadth in terms of usefulness to EA’s diverse teams.

\bigskip

The only directive left is to connect everything together, and this is done in SUPERNOVA with what we call pipelines. They indicate what configured inputs are being used, which model(s) are taking those inputs, and what to do with the output of those models. Possibilities for output actions include simply sending it as input to another model, exporting the data to a specified format, generating email reports, posting the data to an external service, etc. Lastly, the inputs and outputs for each pipeline and its model(s) can be tracked through a built in data visualization tool, which is useful for monitoring of performance over time for longer running automation scripts.

\bigskip

This test selection approach has better fault detection capability than the entire test suite in practice, for the following reasons. First, our game’s can have upwards of 100,000 test cases per title. It is therefore unfeasible for an individual to effectively select which cases to run without a test selection algorithm. Furthermore, if we were to run all the tests with the same distribution, it would become cost and time prohibitive. Second, the number of test cases, as well as the test cases themselves, change over the course of a game's production cycles. A selection system that is unbiased and can quickly be updated based on reactions to the game development workflow is therefore imperative to function effectively in this environment.

\bigskip

Currently we have real world implementations of both the probabilistic RBT modeling and the machine learning. Our deep learning initiatives are still in development, and are intended to expand on these existing models upon completion. We use RBT for automating test case selection, and machine learning for defect prevention on developer commits entering the source code. We will be explaining these initiatives in that order.

\subsection{Use Case \#1: Automated Test Selection with Risk Based Testing}

\begin{table}[h]
\setlength\extrarowheight{5pt}
\begin{tabularx}{\linewidth}{ 
    >{\centering}s 
  | >{\centering}t
  | >{\centering\arraybackslash}b }
  Name & Type & Description \\
 \hline\hline
 Open Unaddressed Defects & Probability & Open unaddressed defects per test case.\\
 \hline
 Addressed Change Requests & Probability & Change requests can impact the game in unknown ways (given that these are requests that are not planned or accounted for). Knowing how many issues will be assessed will give us an indication of risk. \\
 \hline
 Defect to Change Ratio & Probability & The ratio of reported defects per feature change. \\
 \hline
 Script Failure Rate & Probability & The ratio of the positive and negative classes for each test case TC. If a TC has 3 passed, 4 failed and 3 blocked tasks, then its failure rate will be 70\%.\\
 \hline
 Average Distribution & Impact & The average distribution is an automatically computed measure from game telemetry to understand which game modes were most significant to our players. \\
 \hline
 Average Stickiness & Impact & Stickiness is an automatically computed measure to understand the retention of a game mode. \\
 \hline
 QX Final Target & Impact & Expected quality milestone target, which captures production ready feature quality. \\
 \hline
 QX Target vs Current Target & Time & Semi-automatically weighted metric to assess the target vs expected quality milestone. Weights are applied against returned values. \\
 \hline
 Testing Hours in the Last \textit{T} Days & Time & This is a automatically computed metric aggregating the total elapsed time for the past \textit{T} days per test case (where the date is reflective off the tested-on date). \\
 \hline
 Days Since last tested & Time & Each task will have a tested-on date, which is subtracted from the current date to calculate the number of days since lasted tested. \\
 \hline
 Dev Changes in the last \textit{T} days & Probability $\times$ Time & The amount of changes developers make will influence the risk and where we test. As changes increase in a given period, so will the risk associated with a particular task. We put a window on it to keep it timely and relevant.\\
\hline\hline
\end{tabularx}
\caption{Example criteria for a RBT model. These metrics capture coverage and/or yield across many different categories, which allow for robust RBT testing.}
\label{table:1}
\end{table}

The testing scope of video games is growing dramatically and legacy QA approaches are quickly becoming unmanageable and inefficient, even considering a high automation coverage of the product. QA testers are expected to make decisions on where to test the software. With RBT, we can ensure those decisions are data driven ones. When we challenge our assumptions, we gain insights into potential risk. By continuously looking back and reviewing our actions, we learn about new possibilities and change our future approach, by adding or removing automatic criteria and further tuning the manual business driven criteria.

\bigskip

\begin{figure*}[h]
\centering
\includegraphics[width=0.8\textwidth]{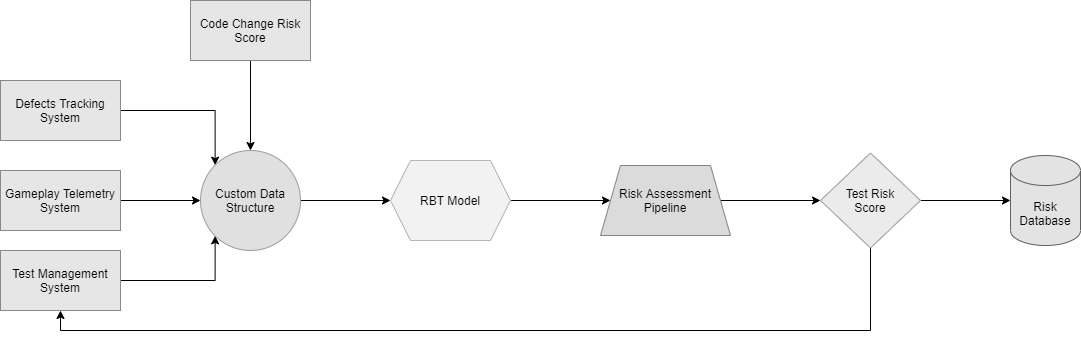}
\caption{Example of the RBT workflow}
\label{rbt}
\end{figure*}

In order to identify a set of RBT metrics, we conduct internal studies on various RBT approaches. Through this analysis and practical experimentation, our QA teams identify the metrics which best describe risky test areas. These metrics are configured via the visual programming interface for creating mathematical models, with examples of them being found in Table \ref{table:1}. They are then fed into an RBT model, where the output of that model is processed and applied to update test case selection, as well as the historical database of risk outputs. A visual representation of this system can be seen in Figure \ref{rbt}.

\subsection{Use Case \#2: Defect Prevention with Machine Learning}

\begin{figure}[h]
\centering
\includegraphics[width=0.3\textwidth]{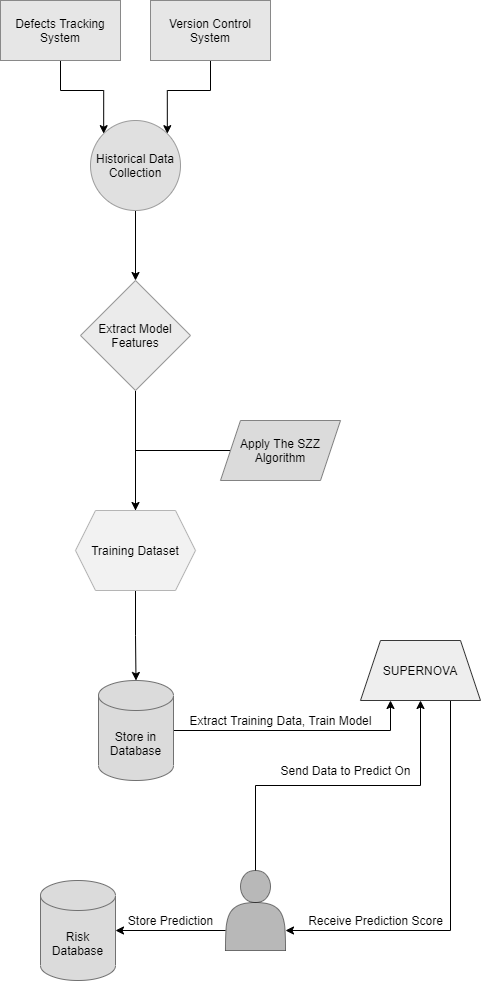}
\caption{Example of the defect prevention workflow}
\label{defectprevention}
\end{figure}

Defects during game development are an inevitable cost but one that can be mitigated by preventing the defect causing code or content to be committed. Machine learning can be used to automatically predict potentially risky code commits based on prior data of known bug inducing commits from bug tracking systems and commit history. However, it takes a lot of effort to create a training data set by manually labelling every commit as risky or non-risky. Hence, semi-supervised machine learning is appropriate for this task. With semi-supervised learning, a heuristic can be used to label the data. We used the SZZ algorithm as the heuristic to label commits as bug inducing. Once the training data set was established, we used a machine learning model to build a binary classifier. The probability of the associated binary classes was then used as the risk score. A diagram showing these relationships can be seen in Figure \ref{defectprevention}.

\bigskip

The risk score prediction is based on a semi-supervised machine learning approach using a tree-based gradient boosting machine (GBM). The gradient boosting framework uses highly optimized tree-based learning which leverages weak learners. Essentially, many iterations of decision tree models are utilized in a gradual, additive and sequential manner, with the aim of reducing the overall error between the true versus predicted class. The process is semi-supervised because the target binary class, whether a commit is bug inducing, is computed using heuristics rather than a manual labelling effort. Ultimately, the larger predicted probability of the target class is assigned as the risk score.

\bigskip

The heuristic used for this semi-supervised learning approach is the base SZZ algorithm, which identifies bug-inducing code commits based on a bug tracking system's historical data (e.g. Jira) and the version control system's commit properties (e.g. Perforce). First, bug reports in the issue tracker are linked to bug-fixing commits. For each of the bug-fixing commits that were identified, the modified lines in the source code are extracted. Second, each bug-fixing commit from the first phase is recursively parsed (e.g. perforce blame) to identify all commits that previously made changes to the same lines of code. For each candidate commit, SZZ determines whether the commit can be ruled out as bug-introducing or not using characteristics like commit date/time, if it was a partial fix, and if it was responsible for another bug. 

\bigskip

However, the accuracy of SZZ isn't perfect, as it has a number of shortcomings. Mills et al. discovered that approximately 64\% of changes made in bug fixing commits are not actually bug fixing, but instead are unrelated changes, such as code additions, refactoring, formatting and comment changes \cite{szzeval1}. SZZ accounts for the additions, but the base algorithm doesn't handle the other cases, and thus approximately 34\% of files labelled as bug inducing commits are false positives. Furthermore, Herzig et al. found that approximately 33\% of bug issue reports are incorrectly categorized, resulting in 39\% of files found to be bug inducing by SZZ being incorrectly labelled. \cite{szzeval2}. These results were independently verified by Herbold et al., and they concluded that "for every correctly labeled defective file, there is one incorrectly labeled defective file and two missed defective files" \cite{szzeval3}. Rosa et al. were among the first to do a comprehensive empirical evaluation of SZZ and all its variants in a controlled environment \cite{szzeval4}. Using a manually verified data set, they avoided the issue of incorrectly categorized bug reports, and thus found that the base SZZ had a recall of 69\%, precision of 42\%, and an F1 score of 53\%. Out of all the SZZ variants tested, R-SZZ performed the best, with a precision of 57\%, recall of 73\%, and an F1 score of 64\% \cite{r-szz}.

\begin{table*}[h]
\centering
\setlength\extrarowheight{5pt}
\begin{tabularx}{\textwidth}{ 
  | >{\centering}X 
  | >{\centering}X 
  | >{\centering}X 
  | >{\centering\arraybackslash}X | }
 \hline
 nFiles $\Delta$ & nCodeFiles $\Delta$ & nFileTypes & nCodeFileTypes\\
 \hline
 nUniqueDir $\Delta$ & nWorkDir $\Delta$ & nLinesAdded $\Delta$ & nLinesEdited $\Delta$\\
 \hline
 nLinesDeleted $\Delta$ & nLinesModified $\Delta$ & nLinesAddedNewFiles $\Delta$ & nLinesDeletedRemovedFiles $\Delta$\\
 \hline
 nLinesTotal $\Psi$ & entropy & nActionAdd $\Delta$ & nActionEdit $\Delta$\\
 \hline
 nActionDelete $\Delta$ & nActionMoveAdd $\Delta$ & nActionMoveDelete $\Delta$ & nActionBranch $\Delta$\\
 \hline
 nActionIntegrate $\Delta$ & nP4TypeText & nP4TypeUtf & nP4TypeBinary\\
 \hline
 revision $\Theta$ & fileSizeTotal & fileSize $\Theta$ & pathDepth $\Theta$\\
 \hline
 lastModifiedElapsed $\Theta$ & developersTotal & developers $\Theta$ & ageCodeFile $\Theta$\\
 \hline
 ageUser & nTokens $\Theta \Psi$ & nFunctions $\Psi$ & nFunctionParameters $\Psi$\\
 \hline
 nComments $\Psi$ & nImports $\Psi$ & codeComplexity $\Theta \Psi$ & codeComplexityAboveThresholdDiff\\
\hline
\end{tabularx}
\caption{These are the features used in our defect prevention model. Entries with  $\Theta$ are one dimensional arrays, and we therefore took the minimum, maximum, mean and median of each to capture their statistical spread in tabular numeric form (i.e. revisionMin, revisionMax, revisionMean, revisionMedian). Entries with  $\Delta$ had an additional entry created to capture historical data for users to infer individual developer patterns (i.e. nFilesUser). Metrics with $\Psi$ had an additional entry created for the previous state of that metric (i.e. nLinesTotalPrev).}
\label{table:2}
\end{table*}

\bigskip

The model features used are found in Table \ref{table:2}, and fall into the following categories: file properties, code properties, commit action properties, and developer properties. For file properties, the age, count, types, and size were used as features. The file age is a representative value of age taken from every file in a commit. The file count is the number of files in the commit, including non-code vs code files. The number of file types is the count of every file suffix, while the file sizes are  representative values for file sizes in the commit. Also, the number of directories is a count of unique directories touched in the commit. For code properties, the lines touched and entropy were useful features. The number of lines touched provides the count of lines that were edited, added, or deleted, while entropy gives the variability of the number of modifications made to files. Code complexity and function counts help infer higher level code data, while imports track number of dependencies and comment counts measure source code documentation.

\bigskip

For commit action properties, we use whether a given file was an addition, deletion, merge, branch, etc. For developer properties, the number of developers implies the count of developers that touched the files in a commit, and the developer age is the total time developer has been on project, starting from their first commit. In the case of features represented by vectors, such as individual file sizes, we took a statistical spread of them by having the minimum, maximum, mean and medium of that vector as features. User historical information and the previous values for a feature were also helpful in providing context for the current commit.

\bigskip

Ultimately, when a developer commits changes to a codebase, the model predicts the likelihood of said commit causing an error. If the prediction passes a certain threshold (risk acceptance level) it alerts the developer, and provides an explanation of the probability output using SHAP values \cite{shap}. The model’s data and metrics will also be available to use for other QA endeavours.

\section{Results}

\subsection{Improvements from Test Selection}

\begin{figure}[h]
\centering
\includegraphics[width=0.45\textwidth]{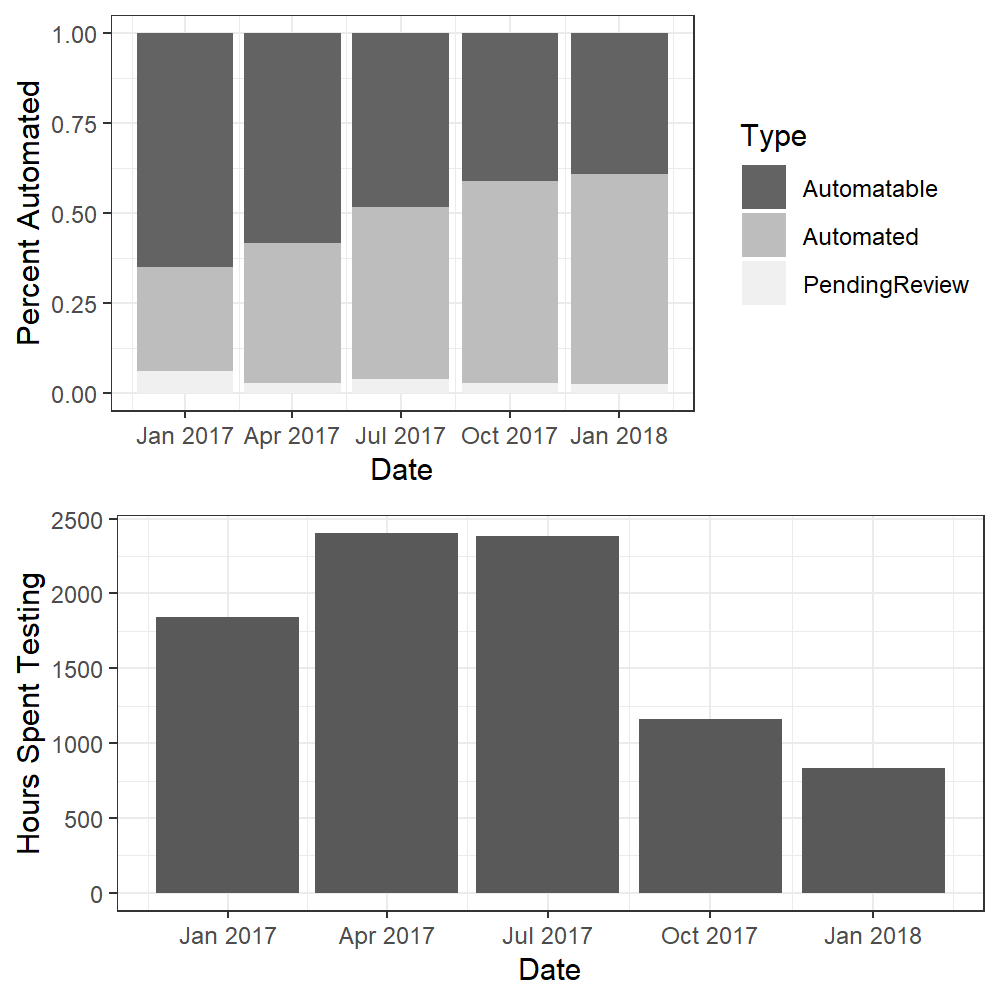}
\caption{These graphs indicate that as test selection automation increased with the adoption of SUPERNOVA, there was a significant drop-off of hours spent testing, as seen in October 2017 and later. The reason that the hour drop-off did not occur earlier is because SUPERNOVA was still a prototype in April and July 2017. Jira tracking was used to determine the hours spent testing for QA teams. Percent automated was determined from assigning a flag to individual test cases, indicating whether or not that test was automated.}
\label{adoptionhours}
\end{figure}

SUPERNOVA has provided improvements to EA's testing teams across several different studios. Looking at Figure \ref{adoptionhours}, as more test selection with RBT became automated, a significant drop off in required testing hours was achieved. From January 2017, before SUPERNOVA was introduced, to one year later, automation increased every quarter from 29\% to 58\%, effectively doubling due to SUPERNOVA. In this time, total hours spent testing also decreased by 55\%. With a quarterly hour decrease of $\sim$1500 hours, that ends up being approximately 6000 less hours spent testing per year. This helps teams to effectively utilize their testing efforts, allowing for unused hours to be reallocated to further improve the quality of our games.

\bigskip

\begin{figure}[h]
\centering
\includegraphics[width=0.5\textwidth]{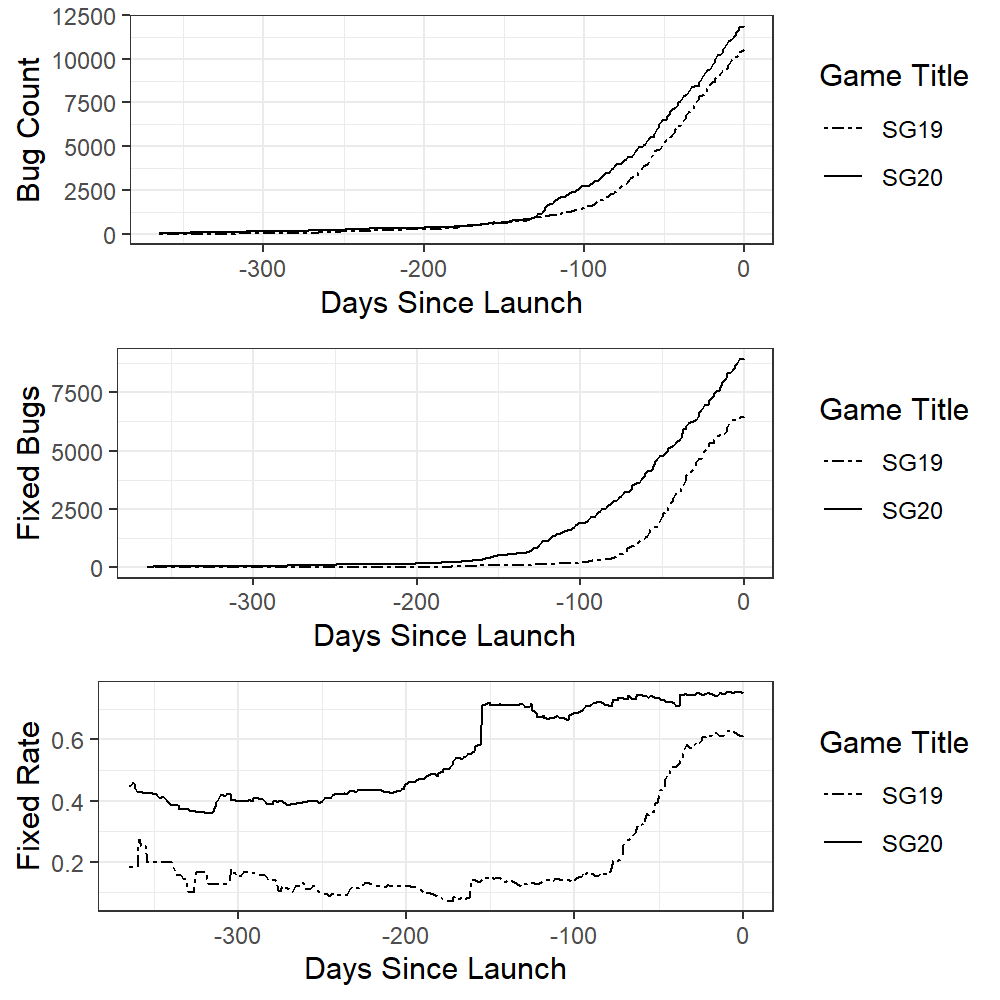}
\caption{Comparison of SUPERNOVA environment with a non-automated approach. SG20 used the automated system, SG19 used manual techniques. Bug count is the cumulative count of bugs found during testing, fixed bugs is the cumulative count of fixed bugs during QA, and fixed rate is fixed bugs / bug count.}
\label{bugs}
\end{figure}

In the year leading up to the launch of two high profile games, sports game 2019 (SG19) and sports game 2020 (SG20), a large difference in QA performance was seen between the two in Figure \ref{bugs}. Efficiency improvements were not just found in working hours. This is because SUPERNOVA was deployed in SG20, but not SG19. The mean daily fix rate of SG20 was 67\% the year leading up to launch. Comparatively, the mean daily fix rate of SG19 was only 25\%. This large difference caused the spike for SG19's fix rate in the three months leading to launch. The spike was driven by crunch time, as the mean fix rate prior to the spike was only 12\%. Meanwhile, SG20 had only a minor crunch spike at 5 months prior to launch, since it's mean fix rate was 45\% prior. Note that there are potentially several hundred individual game testers for a given title, and that testing experience does not dictate testing scope. Therefore, there is no observable correlation between an individual game tester's experience and overall savings in testing hours. This means the higher average fix rate indicates SUPERNOVA significantly reduces crunch time for EA employees to fix bugs. This is a huge quality of life improvement for employees, and reduces the negative impact crunch has on EA studios.

\bigskip

Test planning efforts dropped from 489 hours in SG19 to 0 hours in SG20, with 0.95 bugs found per introduced change versus 0.65 the year prior. The maintenance time of using SUPERNOVA, which has replaced test planning time, was observed to be 12 hours over the period of the fiscal year for SG20. Thus the overall reduction was roughly 97.5\% in test planning overall. There has been an observed on-boarding time of 96 hours with SUPERNOVA for SG20, however this is a one time cost, and not repeated. Accounting for this, the overall test planning savings for SG20 are approximately a 75\% reduction in the first year of development, with further reductions occurring over the course of the game's life cycle, as that initial on-boarding time gets discounted annually.

\bigskip

Additionally, average bugs found per hour reported by QA increased from 0.186 to 0.227, and overall testing time dropped from 48800 hours in SG19 to 9776 hours in SG20. This is because SUPERNOVA helps to reduce test scope by roughly 80\% due to a change towards more focused testing, such as 1100 considered being reduced to 207 selected per testing activity in SG20. The variance of SG20 is also lower overall than SG19, at 0.0075 against 0.0083. During crunch time this difference is further exaggerated, with a variance of 0.0026 against 0.034. This lower variance allows for more accurate projections when planning bug fixing, on top of the already substantially higher fix rate. Furthermore, all of this was accomplished while also finding more bugs to fix. SG19 found 10577 bugs compared to SG20's 11909. So despite finding more bugs, SG20 had a significantly higher bug fix rate than SG19, and notably reduced variance, with SUPERNOVA being the driving factor of these changes.

\subsection{Improvements from Defect Prevention}

SUPERNOVA's defect prevention model was tested on an unreleased game title during its development phase. The results show that the model’s overall macro average performance has 71\% precision, 77\% recall, and an F1 score of 74\%. The minority class performance, which is the positive detection of risky commits, has a 49\% precision, 66\% recall, and an F1 score of 56\%. The majority class performance meanwhile had 94\% precision, 88\% recall, and an F1 score of 91\%. The test set included 1,215 commits labelled as not bug inducing, and 214 commits labelled as bug inducing, since the data set was imbalanced at a ratio of approximately 6:1. These commits flagged as bug inducing represent nearly 20\% of the commits per week which would have otherwise led to additional work hours being put in to fix bugs.

\section{Discussions and Future Work}

\begin{figure}[h]
\centering
\includegraphics[width=0.48\textwidth]{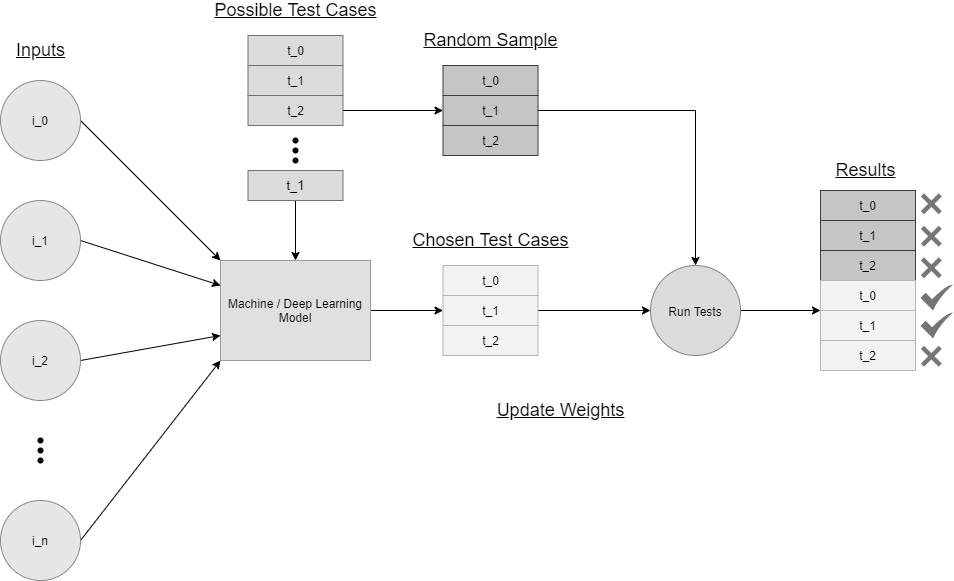}
\caption{Future test selection machine learning architecture}
\label{supernova2}
\end{figure}

In this paper we show that by automating test selection, significant gains can be made across a variety of metrics, such as staff hours, cost, efficiency and consistency. SUPERNOVA is a stepping stone on the path to full testing automation, but for now it should be viewed as a way to transition from old testing methods to new. Future work will include generation of test cases, as well as machine or deep learning models to replace the current RBT and machine learning implementations. Figure \ref{supernova2} shows what deep learning would look like for the test selection process. Essentially, the human factor for weight determination is eliminated, therefore making the process fully automated. To do so, there are two supervised learning options. The first is to simply check which test cases triggered a bug, and which did not, and use this to update the model weights. The second is to randomly sample test cases along with the model output test cases, and check accuracy of both in relation to bugs triggered. This allows for flexibility depending on if test selection needs to be done by the algorithm or not.

\bigskip

Any suitable type of machine or deep learning model may be used to generate and/or refine the test case selection. For example, the test case selection model may be a decision tree, a Gaussian tree, a Bernoulli distribution, a random forest, linear regression, a neural network, a Bayesian network, or any variety of heuristics (e.g. genetic algorithms, swarm algorithms, etc.). These test case selection models could be trained using data that include outcomes from test cases selected in previous rounds of testing (e.g. detection of a bug or not). Finally, an analytic comparison of the test case selection model can be made versus the random selection of test cases.

\bigskip

As for the defect prevention initiative, the next steps involve improving the SZZ algorithm and supplementing the existing machine learning algorithm with a method that captures semantic code properties. For the former, we are planning on adopting the R-SZZ variant, as it achieved notably better performance by ignoring formatting, commenting, and importing changes, while being resilient to refactoring changes. For the latter, this typically involves extracting the abstract syntax trees (ASTs) from a commit's code blocks, and then using an algorithm to perform semantic analysis on these ASTs. This has been demonstrated by CLEVER using clone detection to reduce their model's false positive rate, by comparing suspected bug inducing code with known bug inducing code. 

\bigskip

An alternative method we are investigating is using deep learning to extract features from these ASTs, and then comparing these features with known bug causing features to infer meaning through embeddings \cite{code2vec}. Embeddings are mappings between categorical variables and vectors of continuous numbers, just like encodings, but differ in the context of deep learning. Instead of just static vectors, they are learned representations which are updated as the set of categorical variables change. They also have the benefit of being lower dimensional than their source categorical variables. This is quite useful for tasks involving programming and source code, as the scale is quite large and variance between programming languages is significant.

\section{Conclusion}

The rapidly advancing costs associated with game and software testing have become prohibitive towards brute force methods. Many market solutions have sprung up to solve this, yet none of them are capable of working with test cases as complex as the ones at EA, or are able to fit into our existing pipelines. SUPERNOVA allows for QA testers to substantially reduce the amount of time spent on the planning of testing programs, and divert this to other more productive ventures. Furthermore, it seeks to tackle the other end of the problem as well, by preventing defects from entering the system altogether. It increases efficiencies and is more reliable than traditional methods, allowing for better forecasting. EA is excited to continue moving forward with this innovative approach in games testing, to keep giving our players the best experiences.

\nocite{*}
\bibliographystyle{ieeetr}
\bibliography{references}

\begin{thebibliography}{10}

\bibitem{Munro2009}
J.~Munro, C.~Boldyreff, and A.~Capiluppi, ``{Architectural Studies of Games
  Engines — The Quake Series},'' in {\em 2009 International IEEE Consumer
  Electronics Society's Games Innovations Conference}, pp.~246--255, 2009.

\bibitem{patent}
M.~Culibrk, D.~Ispir, and A.~Senchenko, ``{Optimized Test Case Selection For
  Quality Assurance Testing Of Video Games},'' 2020.
\newblock US 20200310948 A1.

\bibitem{katalon}
Z.~Ereiz {\em et~al.}, ``{Automating Web Application Testing Using Katalon
  Studio},'' {\em Zbornik Radova Medunarodne Nau{\v{c}}ne Konferencije o
  Digitalnoj Ekonomiji DIEC}, vol.~2, no.~2, pp.~87--97, 2019.

\bibitem{selenium}
A.~Bruns, A.~Kornstadt, and D.~Wichmann, ``{Web Application Tests with
  Selenium},'' {\em IEEE software}, vol.~26, no.~5, pp.~88--91, 2009.

\bibitem{selenium2}
P.~Ramya, V.~Sindhura, and P.~V. Sagar, ``{Testing Using Selenium Web
  Driver},'' in {\em 2017 Second International Conference on Electrical,
  Computer and Communication Technologies (ICECCT)}, pp.~1--7, IEEE, 2017.

\bibitem{testcomplete}
O.~Shakurova, ``{Automating UI Tests for a Web Application Using
  Test-Complete},'' 2015.
\newblock BSc Thesis HAAGA-HELIA Ammattikorkeakoulu.

\bibitem{testcomplete2}
S.~Al-Zain, D.~Eleyan, and J.~Garfield, ``{Automated User Interface Testing for
  Web Applications and TestComplete},'' in {\em Proceedings of the CUBE
  International Information Technology Conference}, pp.~350--354, 2012.

\bibitem{dataflow}
S.~Rapps and E.~J. Weyuker, ``{Selecting Software Test Data using Data Flow
  Information},'' {\em IEEE Transactions on Software Engineering}, no.~4,
  pp.~367--375, 1985.

\bibitem{testselection1}
M.~Gligoric, L.~Eloussi, and D.~Marinov, ``Practical regression test selection
  with dynamic file dependencies,'' in {\em Proceedings of the 2015
  International Symposium on Software Testing and Analysis}, ISSTA 2015, (New
  York, NY, USA), p.~211–222, Association for Computing Machinery, 2015.

\bibitem{testselection2}
O.~Legunsen, F.~Hariri, A.~Shi, Y.~Lu, L.~Zhang, and D.~Marinov, ``An extensive
  study of static regression test selection in modern software evolution,'' FSE
  2016, (New York, NY, USA), p.~583–594, Association for Computing Machinery,
  2016.

\bibitem{testselection3}
L.~Zhang, ``Hybrid regression test selection,'' ICSE '18, (New York, NY, USA),
  p.~199–209, Association for Computing Machinery, 2018.

\bibitem{infer}
D.~Distefano, M.~F{\"a}hndrich, F.~Logozzo, and P.~W. O'Hearn, ``{Scaling
  Static Analyses at Facebook},'' {\em Communications of the ACM}, vol.~62,
  no.~8, pp.~62--70, 2019.

\bibitem{clevercommit}
M.~Nayrolles and A.~Hamou-Lhadj, ``{CLEVER: Combining Code Metrics with Clone
  Detection for Just-in-Time Fault Prevention and Resolution in Large
  Industrial Projects},'' in {\em Proceedings of the 15th International
  Conference on Mining Software Repositories (MSR)}, pp.~153--164, 2018.

\bibitem{szz}
J.~{\'S}liwerski, T.~Zimmermann, and A.~Zeller, ``{When Do Changes Induce
  Fixes?},'' {\em ACM SIGSOFT Software Engineering Notes}, vol.~30, no.~4,
  pp.~1--5, 2005.

\bibitem{Kondo2020}
M.~Kondo, D.~M. German, O.~Mizuno, and E.-H. Choi, ``{The Impact of Context
  Metrics on Just-in-Time Defect Prediction},'' {\em Empirical Software
  Engineering}, vol.~25, no.~1, pp.~890--939, 2020.

\bibitem{Tabassum2020}
S.~Tabassum, L.~L. Minku, D.~Feng, G.~G. Cabral, and L.~Song, ``{An
  Investigation of Cross-Project Learning in Online Just-in-Time Software
  Defect Prediction},'' in {\em 2020 IEEE/ACM 42nd International Conference on
  Software Engineering (ICSE)}, pp.~554--565, IEEE, 2020.

\bibitem{Chen2014}
T.-H. Chen, M.~Nagappan, E.~Shihab, and A.~E. Hassan, ``{An Empirical Study of
  Dormant Bugs},'' in {\em Proceedings of the 11th Working Conference on Mining
  Software Repositories (MSR)}, pp.~82--91, 2014.

\bibitem{rbt}
M.~Felderer, C.~Haisjackl, V.~Pekar, and R.~Breu, ``An exploratory study on
  risk estimation in risk-based testing approaches,'' vol.~200, 01 2016.

\bibitem{rbt2}
M.~Felderer, C.~Haisjackl, R.~Breu, and J.~Motz, ``Integrating manual and
  automatic risk assessment for risk-based testing,'' vol.~94, pp.~159--180, 01
  2012.

\bibitem{versionctrl}
N.~N. Zolkifli, A.~Ngah, and A.~Deraman, ``{Version Control System: A
  Review},'' {\em Procedia Computer Science}, vol.~135, pp.~408--415, 2018.

\bibitem{szzeval1}
C.~Mills, J.~Pantiuchina, E.~Parra, G.~Bavota, and S.~Haiduc, ``Are bug reports
  enough for text retrieval-based bug localization?,'' pp.~381--392, 09 2018.

\bibitem{szzeval2}
K.~Herzig, S.~Just, and A.~Zeller, ``It's not a bug, it's a feature: How
  misclassification impacts bug prediction,'' in {\em Proceedings of the 2013
  International Conference on Software Engineering}, ICSE '13, p.~392–401,
  IEEE Press, 2013.

\bibitem{szzeval3}
S.~Herbold, A.~Trautsch, F.~Trautsch, and B.~Ledel, ``Problems with szz and
  features: An empirical study of the state of practice of defect prediction
  data collection,'' {\em Empirical Software Engineering}, vol.~27, 03 2022.

\bibitem{szzeval4}
G.~Rosa, L.~Pascarella, S.~Scalabrino, R.~Tufano, G.~Bavota, M.~Lanza, and
  R.~Oliveto, {\em Evaluating SZZ Implementations Through a Developer-Informed
  Oracle}, p.~436–447.
\newblock IEEE Press, 2021.

\bibitem{r-szz}
S.~Davies, M.~Roper, and M.~Wood, ``Comparing text-based and dependence-based
  approaches for determining the origins of bugs,'' {\em Journal of Software:
  Evolution and Process}, vol.~26, 01 2014.

\bibitem{shap}
S.~M. Lundberg, G.~G. Erion, and S.-I. Lee, ``Consistent individualized feature
  attribution for tree ensembles,'' {\em arXiv preprint arXiv:1802.03888},
  2018.

\bibitem{code2vec}
U.~Alon, M.~Zilberstein, O.~Levy, and E.~Yahav, ``Code2vec: Learning
  distributed representations of code,'' {\em Proc. ACM Program. Lang.},
  vol.~3, jan 2019.

\end{thebibliography}

\end{document}